\documentstyle[12pt]{article}
\newcommand{\beq}{\begin{equation}}
\newcommand{\eeq}{\end{equation}}
\newcommand{\beqa}{\begin{eqnarray}}
\newcommand{\eeqa}{\end{eqnarray}}
\newcommand{\ba}{\begin{array}}
\newcommand{\ea}{\end{array}}

\begin{document}

\begin{center}
{\large \bf Spectral Statistics in Large Shell Model Calculations}
\footnote{This work has been partially supported by the Ministero
dell'Universit\`a e della Ricerca Scientifica e Tecnologica (MURST)
and by the INFN--CICYT Agreement.}
\end{center}

\vskip 0.5 truecm

\begin{center}
{\bf J.M.G. G\'omez}
\vskip 0.5 truecm
Departamento de Fisica At\'omica, Molecular y Nuclear, \\
Facultad de Ciencias Fisicas, Universidad Complutense de Madrid, \\
E--28040 Madrid, Spain \\

\vskip 0.5 truecm

{\bf V.R. Manfredi} \\
\vskip 0.5 truecm
Dipartimento di Fisica "G. Galilei" dell'Universit\`a di Padova, \\
INFN, Sezione di Padova, \\
Via Marzolo 8, I--35131 Padova, Italy \\
Interdisciplinary Laboratory, International School for Advanced Studies, \\
Strada Costiera 11, I--34014 Trieste, Italy

\vskip 0.5 truecm

{\bf L. Salasnich} \\
\vskip 0.5 truecm
Departamento de F\'isica At\'omica, Molecular y Nuclear, \\
Facultad de Ciencias Fisicas, Universidad Complutense de Madrid, \\
E--28040 Madrid, Spain \\

\end{center}

\vskip 0.8 truecm

\begin{center}
{\bf Abstract}
\end{center}
\vskip 0.5 truecm
\par
The spectral statistics of low--lying states of $fp$ shell nuclei are studied
by performing large shell--model calculations with a realistic nuclear
interaction. For $Ca$ isotopes, we find deviations from the predictions of the
random--matrix theory which suggest that some spherical nuclei
are not as chaotic in nature as the conventional view assumes.

\vskip 1.5 truecm

\par
{\bf 1. Introduction}
\vskip 0.5 truecm
\par
In recent years many authors have shown great interest in the fluctuation
properties of energy levels$^{1)}$. It is well known that the fluctuation
properties of quantum systems with underlying classical chaotic behaviour and
time--reversal symmetry agree with the predictions of the Gaussian
Orthogonal Ensemble (GOE) of the random--matrix theory, whereas
quantum analogs of classically integrable systems display the characteristics
of the Poisson statistic$^{2)}$.
\par
In atomic nuclei, the fluctuation properties of energy levels are
best studied in the domain of neutron and proton resonances near the nucleon
emission threshold, where a large number of levels of the same spin and parity
in the same nucleus are present,
and an excellent agreement with GOE predictions
has been found$^{3)}$. In the ground--state region, however, the samples of
consecutive experimental levels of the same spin and parity in any one
nucleus are quite small. Therefore it is more difficult to calculate
reliable mean values and fluctuations of statistics such as energy level
spacings. In order to circumvent this difficulty, in recent years$^{4,5,6)}$
statistical analyses of experimental lower excitation energies have
combined data from a large range of excitation energies and angular
momenta of a nucleus or a set of nuclei. Such analyses have provided
evidence suggesting that spherical nuclei show level spectra close to GOE
predictions and deformed nuclei show strong deviations from GOE behaviour.
In a recent analysis$^{7)}$ of level spacings close to the yrast line of
deformed nuclei with $Z=62$--$75$ and $A=155$--$185$,
the average level spacing for
states with the same spin and parity was calculated for the total ensemble
instead of for individual nuclei. The level spacing fluctuations
obtained are quite close to the Poisson distribution, showing evidence of
regular motion.
\par
We can conclude that experimental nuclear spectra suggest a dominance
of chaotic motion in most cases and that regular motion seems to be
prevalent in a small region of excitation energy just above the yrast line
of deformed nuclei. But the borderlines in mass number, excitation energy,
etc., between order and chaos in nuclei are not sharply distinguished
by the experimental data available.

\vskip 0.5 truecm
\par
{\bf 2. Analysis of the fp shell}
\vskip 0.5 truecm
\par
\par
Theoretical models should also help to establish the domains of chaos in
nuclei. First of all, the success of the nuclear shell model and the
collective models of nuclear structure at low excitation energies provide
strong evidence of regular motion in nuclei.
Theoretical studies of deformed nuclei have shown that a two--body
interaction of reasonable strength causes chaotic spectra a few hundred keV
above the yrast line due to rotational band mixing$^{8)}$.
The transition from order to chaos has also been studied in the framework of
the Interacting Boson Model and was found to depend on the
values of the model parameters. Regular features of the spectra were
found to be associated with values of the parameters for which
the hamiltonian displays a symmetry$^{9)}$.
However, statistical analyses of shell model energy
spectra and wave functions have almost always shown that
chaos in nuclei is quite widespread$^{10)}$.
\par
It is somewhat surprising that shell model calculations suggest such a
prevalence of chaos because the model itself is based on a mean field
idea associated with regular motion. The residual interaction can
destroy the regularity of single--particle orbits and cause the chaotic
features, but it is still remarkable that chaos can be so dominant
even for the lowest energy levels of light nuclei$^{11)}$.
Recent shell--model calculations by Bae et al.$^{12)}$ have shown that for
nuclei of mass $A=212$ shell model spectra display
features of regular motion in some cases. The main reason
for this seems to be the relatively small values of the residual
interaction matrix elements as compared to the average spacing between
neighbouring single--particle levels in heavy nuclei.

\vskip 0.5 truecm
{Table 1: m--scheme and maximal (J,T) dimensions of the
analyzed configurations (n,p) of active nucleons}.
\vskip 0.5 truecm
\begin{center}
\begin{tabular}{|ccccccc|} \hline\hline
{}~~ & $^{46}V$ & $^{46}Ti$ & $^{46}Sc$ & $^{46}Ca$ & $^{48}Ca$ & $^{50}Ca$
\\ \hline
$(n,p)$ & (3,3) & (4,2) & (5,1) & (6,0) & (8,0) & (10,0) \\ \hline
m--scheme dimension
& $121\; 440$ & $86\; 810$ & $30\; 042$ & $3\; 952$ & $12\; 002$ &
$17\; 276$ \\
$(J,T)$ & $(4,0)$ & $(4,1)$ & $(4,2)$ & $(4,3)$ & (4,4) & (4,5) \\
$(J,T)$ dimension & $4\; 750$ & $8\; 026$ & $3\; 783$ & $615$ & $1\; 755$
& $2\; 468$
\\ \hline\hline
\end{tabular}
\end{center}

\vskip 0.5 truecm

\par
In this work, we undertake the statistical analysis of the shell--model
energy levels in the $A=46$--$50$ region. Exact calculations
are performed in the ($p_{1/2}$,$p_{3/2}$,$f_{5/2}$,$f_{7/2}$)
shell--model space, assuming a $^{40}Ca$ inert core.
The diagonalizations are performed in the {\it m--scheme} using a fast
implementation of the Lanczos algorithm through the code ANTOINE$^{14)}$.
The interaction we use is a minimally modified Kuo--Brown
force, explained in$^{15)}$. For a fixed number of
valence protons and neutrons we calculate the energy spectrum for
projected total angular momentum $J$ and total isospin $T$.

\vskip 0.5 truecm
{Table 2: Number of energy levels up to $4$ and $5$ MeV
above the yrast line in $^{46}V$}.
\vskip 0.5 truecm
\begin{center}
\begin{tabular}{|ccc|} \hline\hline
$J$ & $4$ MeV & $5$ MeV \\ \hline
$0$ & $1$ & $2$ \\
$1$ & $7$ & $11$ \\
$2$ & $5$ & $8$ \\
$3$ & $14$ & $23$ \\
$4$ & $7$ & $11$ \\
$5$ & $13$ & $23$ \\
$6$ & $8$ & $13$ \\
$7$ & $18$ & $33$ \\
$8$ & $9$ & $14$ \\
$9$ & $14$ & $21$ \\ \hline
$0$--$9$ & $96$ & $159$ \\ \hline\hline
\end{tabular}
\end{center}

\vskip 0.5 truecm

\par
We calculate the $T=T_z$ states from $J=0$ to $J=9$
for all the combinations $(n,p)$ of $6$ active nucleons,
i.e. (3,3), (4,2), (5,1) and (6,0). For the sake of completeness
we also calculated (8,0) and (10,0).
Table 1 shows
the m--scheme and the maximal (J,T) dimensions of the
configurations $(n,p)$ of active nucleons analyzed.

\par
Since we are looking for deviations from chaotic features, we are
mainly interested in the low--lying levels, up to a few MeV above the $JT$
yrast line. Table 2 shows the number of levels up to $4$ and $5$ MeV
above the yrast line in $^{46}V$. As can be seen, in many cases
the number of levels is too small to calculate fluctuations
around the average spacing between neighbouring levels.

\par
Hence in those cases we use a larger set of levels for the
calculation of the mean level spacing as a function of energy.
For each $JT$ set of levels the spectrum is
mapped into unfolded levels with quasi--uniform level density
by using the constant temperature formula$^{5)}$. Under the $8$ MeV range
of energies the mean level density can be assumed to be of the form
\beq
{\bar \rho}(E)={1\over T}\exp{[(E-E_0)/T]} ,
\eeq
where $T$ and $E_0$ are constants. For fitting purposes it is better
to use not ${\bar \rho}(E)$ but its integral ${\bar N}(E)$. We take
\beq
{\bar N}(E)=\int_0^E {\bar \rho}(E')dE' + N_0 = \exp{[(E-E_0)/T]}-
\exp{[-E_0/T]}+N_0 .
\eeq
The constant $N_0$ represents the number of levels with energies less
than zero. Following Shriner et al.$^{5)}$, we consider Eq. 2 as an
empirical function to fit the data and let $N_0$ take non--zero values.
The parameters $T$, $E_0$ and $N_0$
that best fit $N(E)$ are obtained by minimizing the function:
\beq
G(T,E_0,N_0)=\int_{E_{min}}^{E_{max}} [N(E)-{\bar N}(E)]^2 dE ,
\eeq
where $N(E)$ is the number of levels with energies less than or equal to $E$.
The energies $E_{min}$ and $E_{max}$ are taken as the first and last
energies of the level sequence. As an example, Fig. 1 illustrates
the fit to the integrated level density $N(E)$ for the $J^{\pi}T=6^+1$
levels of $^{46}Ti$.

\begin{figure}
\vskip 9. truecm
\caption{The best fit to the integrated level density for $^{46}Ti$
with $J=6^+$ and isospin $T-T_z =0$.}
\vskip 0.5 truecm
\end{figure}

\par
Once the best fit $F(E)$ to $N(E)$ is obtained, the unfolded energy
levels are given by
\beq
{\tilde E_i}=E_{min}+{F(E_i)-F(E_{min})\over
F(E_{max})-F(E_{min})}(E_{max}-E_{min}) .
\eeq
These transformed energies should now display on average a constant
level density $\rho_c$.
\par
The spectral statistic $P(s)$ is used
to study the local fluctuations of the energy levels$^{16,17)}$. $P(s)$ is
the distribution of nearest--neighbour spacings
$s_i=({\tilde E}_{i+1}-{\tilde E}_i)\rho_c$ of the unfolded levels ${\tilde
E}_i$. It is obtained by accumulating the number of spacings that lie within
the bin $(s,s+\Delta s)$ and then normalizing $P(s)$ to unity.
\par
For quantum systems whose classical analogs are integrable,
$P(s)$ is expected to follow the Poisson limit, i.e.
$P(s)=\exp{(-s)}$. On the other hand,
quantal analogs of chaotic systems exhibit the spectral properties of
GOE with $P(s)= (\pi / 2) s \exp{(-{\pi \over 4}s^2)}$ $^{1,2)}$.
\par
The $P(s)$ distribution is compared to the Brody distribution$^{18)}$
\beq
P(s,\omega)=\alpha (\omega +1) s^{\omega} \exp{(-\alpha s^{\omega+1})},
\eeq
with
\beq
\alpha = (\Gamma [{\omega +2\over \omega+1}])^{\omega +1}.
\eeq
This distribution interpolates between the Poisson distribution ($\omega =0$)
of integrable systems and the GOE distribution ($\omega =1$) of
chaotic ones.
\par
In order to obtain a meaningful statistic, $P(s)$ is calculated using the
unfolded level spacings of the whole set of $J=0$--$9$ levels
for fixed $T$ up to a given energy limit above the yrast line.
Thus the number of spacings included is reasonably large. For
example, up to $4$ MeV there are $52$ spacings for $T=0$ and $86$
for $T=1$, and up to $5$ MeV there are $87$ and $149$ spacings,
respectively.
Table 3 shows the Brody parameter $\omega$ for the $J=0$--$9$ set of
level spacings for the $A=46$ nuclei up to $4$ and $5$ MeV above
the yrast line.

\par
Clearly, $^{46}V$ and $^{46}Ti$ are highy chaotic,
but there is a considerable deviation from GOE
predictions in $^{46}Ca$, which is a single closed shell nucleus.
Similar calculations for $^{48}Ca$ and $^{50}Ca$ yield $\omega$ values
similar to $^{46}Ca$, as shown in Table 3. Thus, for the $Ca$
isotopes we find the same kind of phenomenon obtained by Bae et al.$^{12)}$
in the heavy single closed nuclei $^{212}Rn$ and $^{212}Pb$,
namely that low--lying states deviate strongly from chaoticity
toward regularity.

\begin{figure}
\vskip 9. truecm
\caption{$P(s)$ for low--lying levels of the $fp$ shell with
$0\leq J \leq 9$ for $(n,p)$ configurations: $(3,3)$ and $(4,2)$.
The dotted, dashed and solid curves stand for GOE,
Poisson, and Brody distributions, respectively.}
\vskip 0.5 truecm
\end{figure}

\vskip 0.5 truecm

{Table 3: Brody parameter $\omega$ for the fp shell for different
configurations (n,p) of active nucleons with $0\leq J\leq 9$
and $E\leq 4$ MeV (up) and $E\leq 5$ MeV (down)}.
\vskip 0.5 truecm
\begin{center}
\begin{tabular}{|cccccc|} \hline\hline
$^{46}V$ & $^{46}Ti$ & $^{46}Sc$ & $^{46}Ca$ & $^{48}Ca$ & $^{50}Ca$
\\ \hline
0.97 & 1.09 & 0.65 & 0.57 & 0.64 & 0.56 \\
0.97 & 0.97 & 0.93 & 0.57 & 0.50 & 0.65 \\ \hline\hline
\end{tabular}
\end{center}

\vskip 0.5 truecm

\par
To obtain a better estimate of the Brody parameter, we can separately combine
spacings of different nuclei. In fig. 2 we plot $P(s)$ for the
$236$ spacings of $^{46}V$+$^{46}Ti$, and in Fig. 3 for the $266$ spacings
of $^{46}Ca$+$^{48}Ca$+$^{50}Ca$ up to $5$ MeV above the yrast lines.
The number of level spacings is now sufficiently large to yield
meaningful statistics and we see that $Ca$ isotopes are not very
chaotic at low energy, in contrast to other nuclei in the same region.
\par
How may these results be explained? We observe that
the two--body matrix elements of the proton--neutron interaction are,
on average, larger than those of
the proton--proton and neutron--neutron interactions.
Consequently the single--particle motion in nuclei with both
protons and neutrons in the valence orbits suffers
more disturbance and is thus more chaotic.
The shift towards the Poisson distribution obtained for the $Ca$ isotopes
could also be due to some underlying symmetry dominating the
dynamics of those nuclei. However, we do not find significant differences
between $J=0$ and high $J$ values when they are analyzed separately.
Thus the seniority scheme does not seem to have a significant influence
on the shift towards regularity observed in $Ca$ isotopes.

\begin{figure}
\vskip 9. truecm
\caption{$P(s)$ for low--lying levels of the $fp$ shell with
$0\leq J \leq 9$ for $(n,p)$ configurations: $(6,0)$, $(8,0)$ and $(10,0)$.
The dotted, dashed and solid curves stand for GOE,
Poisson, and Brody distributions, respectively.}
\vskip 0.5 truecm
\end{figure}

\par
It should be noted that the analysis$^{19)}$ of experimental energy levels
below $4.3$ MeV excitation energy in the semi--magic nucleus
$^{116}Sn$ yields a near--neighbour spacing distribution which is
intermediate between GOE and Poisson, with
$\omega =0.51 \pm 0.19$. This result is consistent with the theoretical
findings of Bae et al.$^{12)}$ for $^{212}Rn$ and $^{212}Pb$, and
our present results for $Ca$ isotopes.

\vskip 0.5 truecm
\par
{\bf 3. Conclusions}
\vskip 0.5 truecm
\par
Why do all shell model calculations give chaotic features for $sd$ shell
nuclei, without any significant deviations towards regularity? First, it
should be noted that most of these calculations include a large
number of states, up to excitation energies far above the nucleon
emission threshold. This is, for example, the case of the $^{22}O$
calculations of Bae et al.$^{12)}$, which include the full set of
levels for several $J$ values and obtain $\omega =0.96$.
Second, we notice that Ormand and Broglia$^{11)}$
obtained a GOE--like distribution
for the first two spacings of each spectrum for a set of $sd$ shell
nuclei. However,
these nuclei have valence protons and neutrons and are
thus similar to the case of $^{46}V$ and $^{46}Ti$, for which we also
find chaotic behaviour. We conclude that no regular features have been found
in the $sd$ shell region. This is because single--closed nuclei have a small
configuration space and too few low--lying levels for statistical
analysis, and also because the disturbance of single--particle motion
by the two--body interaction is greatest in light nuclei.

\par

\vskip 0.5 truecm

\parindent=0. pt
\section*{References}

\vskip 0.5 truecm

1. O. Bohigas, M. J. Giannoni, and C. Schmit, in {\it Quantum Chaos and
Statistical Nuclear Physics}, Ed. T. H. Seligman and H. Nishioka
(Springer--Verlag, Berlin, 1986);
O. Bohigas and H. A. Weidenm\"uller, Ann. Rev. Nucl. Part. Sci.
{\bf 38}, 421 (1988)

2. M. C. Gutzwiller, {\it Chaos in Classical and Quantum Mechanics}
(Springer--Verlag, Berlin, 1990);
A. M. Ozorio de Almeida, {\it Hamiltonian Systems: Chaos and
Quantization} (Cambridge University Press, Cambridge, 1990);
K. Nakamura, {\it Quantum Chaos} (Cambridge Nonlinear Science Series,
Cambridge, 1993);
{\it From Classical to Quantum Chaos}, SIF Conference Proceedings, vol.
{\bf 41}, Ed. G. F. Dell'Antonio, S.
Fantoni, V. R. Manfredi (Editrice Compositori, Bologna, 1993)

3. R. U. Haq, A. Pandey, and O. Bohigas, Phys. Rev. Lett. {\bf 48}, 1086
(1982)

4. G. E. Mitchell, E. G. Bilpuch, P. M. Endt, and J. F. Jr. Shriner, Phys. Rev.
Lett. {\bf 61}, 1473 (1988);

5. J. F. Jr. Shriner, E. G. Bilpuch, P. M. Endt, and
G. E. Mitchell, Z. Phys. A {\bf 335}, 393 (1990);
J. F. Jr. Shriner, G. E. Mitchel, and T. von Egidy, Z. Phys. A {\bf 338},
309 (1990)

6. J. F. Jr. Shriner, G. E. Mitchell, and T. von Egidy,
Z. Phys. {\bf 338}, 309 (1991)

7. J. D. Garrett, J. R. German, L. Courtney, and J. M. Espino, in
{\it Future Directions in Nuclear Physics}, p. 345,
Ed. J. Dudek and B. Haas (American Institute of Physics,
New York, 1992);
M. T. Lopez--Arias, V. R. Manfredi, and L. Salasnich, Riv. Nuovo Cim.
{\bf 17}, N. 5, 1 (1994)

8. S. Aberg, Phys. Rev. Lett. {\bf 64}, 3119 (1990)

9. N. Whelan and Y. Alhassid, Nucl. Phys. A {\bf 556}, 42 (1993)

10. T. A. Brody, J. Flores, J. B. French, P. A. Mello, A. Pandey,
and S. S. Wong, Rev. Mod. Phys. {\bf 53}, 385 (1981);
H. Dias, M. S. Hussein, N. A. de Oliveira, and B. H. Wildenthal,
J. Phys. G {\bf 15}, L79 (1989); V. Paar, D. Vorkapic, K. Heyde,
A. G. M. van Hees, and A. A. Wolters, Phys. Lett. B {\bf 271}, 1 (1991)

11. W. E. Ormand and R.A. Broglia, Phys. Rev. C {\bf 46}, 1710 (1992)

12. M. S. Bae, T. Otshka, T. Mizusuki, and N. Fukumishi, Phys. Rev. Lett.
{\bf 69}, 2349 (1992)

13. R. D. Lawson, {\it Theory of the Nuclear Shell Model}
(Clarendon, Oxford 1980)

14. E. Caurier, computer code ANTOINE, CRN, Strasburg (1989);
E. Caurier, A. P. Zuker, and A. Poves: in {\it Nuclear Structure of
Light Nuclei far from Stability. Experiment ad Theory}, Proceedings of
the Workshop, Obrnai, Ed. G. Klotz (CRN, Strasburg, 1989)

15. E. Caurier, A. P. Zuker, A. Poves, and G. Martinez--Pinedo,
Phys. Rev. C {\bf 50}, 225 (1994)

16. F. J. Dyson and M. L. Mehta, J. Math. Phys. {\bf 4}, 701 (1963)

17. O. Bohigas and M. J. Giannoni, Ann. Phys. (N.Y.) {\bf 89}, 393 (1975)

18. T. A. Brody, Lett. Nuovo Cimento {\bf 7}, 482 (1973)

19. S. Raman et al., Phys. Rev. C {\bf 43}, 521 (1991)

\end{document}